\documentclass[preprint,amsmath,amssymb,a4paper,prb,superscriptaddress]{revtex4}
\usepackage[dvipdfm]{graphicx}
\usepackage{natbib}
\usepackage{multirow}
\usepackage{amsmath}
\usepackage{url}
\usepackage{mathrsfs}

\begin{document}
\title{ Phonon-phonon interactions in transition metals }

\author{Laurent Chaput} \affiliation{Institut Jean Lamour, UMR CNRS 7198, Nancy Universit\'{e}, Bd. des Aiguillettes, BP 23, 54506 Vandoeuvre Les Nancy Cedex, France}
\email{laurent.chaput@ijl.nancy-universite.fr}
\author{Atsushi Togo}
\affiliation{ Laboratoire d'Etude des Microstructures, UMR 104 ONERA-CNRS
ONERA, BP 72, 92322 Ch\^atillon cedex, France }
\affiliation{ Department of Materials Science and
Engineering, Kyoto University, Sakyo, Kyoto 606-8501, Japan }

\author{Isao Tanaka} \affiliation{ Department of Materials Science and
Engineering, Kyoto University, Sakyo, Kyoto 606-8501, Japan }
\affiliation{ Nanostructures Research Laboratory, Japan Fine Ceramics
Center, Atsuta, Nagoya 456-8587, Japan}

\author{Gilles Hug} \affiliation{ Laboratoire d'Etude des
Microstructures, UMR 104 ONERA-CNRS ONERA, BP 72, 92322 Ch\^atillon
cedex, France }

\begin{abstract}
In this paper the phonon self energy produced by anharmonicity is calculated using second order many body perturbation theory for all bcc, fcc and hcp transition metals. The symmetry properties of the phonon interactions are used to obtain an expression for the self energy as a sum over irreducible triplets, very similar to integration in the irreducible part of the Brillouin zone for one particle properties. The results obtained for transition metals shows that the lifetime is on the order of $10^{-10}$s. Moreover the Peierls approximation for the imaginary part of the self energy is shown to be reasonable for bcc and fcc metals. For hcp metals we show that the Raman active mode decays into a pair of acoustic phonons, their wave vector being located on a surface defined by conservation laws.\\ 
\end{abstract}

\maketitle

Harmonic phonon calculations based on density functional theory are nowadays routinely performed for bulk solids. The dynamical matrix is either obtained from density functional perturbation theory\cite{baroni-1987} or from supercell calculations\cite{parlinski-1997,phonopy}. To go beyond the harmonic approximation quasiharmonic calculations are usually performed\cite{pavone,togo10}. However in this effective theory the phonons do not have lifetime.  \emph{Ab-initio} anharmonic calculations taking into account phonon-phonon interactions explicitly are rather rare. There are noticeable exceptions with for example the calculations in the diamond structure of Si and Ge \cite{deinzer-2003,narasimhan-1991}, and the recent study of graphite by Bonini et \emph{al} \cite{bonini-2007}. Such calculations gives relevant informations about the phonon-phonon interactions which may be hidden by the electron-phonon interaction in experiments.  It is important, for example, in the understandig of energy transport in thermoelectricity. Looking at the self energy of simple basic elements is therefore of interest.

In this paper we study the bcc, fcc and hcp transition metals for the first time. The phonon-phonon self energy is calculated for all metals in the crystallographic structure stable under normal conditions. The necessary information is then extracted to obtain the decay path for a selected phonon in the Brillouin zone.

The paper is organized as follow. To explain the methodology of our calculation we first define irreducible triplets of wave vectors from the symmetry of the phonon-phonon coupling function. A formula is then obtained to calculate the $({q},\omega)$ resolved self energy in a simple way. The results of these calculations for the transition metals are then analyzed using band decomposition and conservation surfaces for the phonons with the shorter lifetime. An approximation proposed by Peierls is also discussed. Finally the phonon decay path generating the Raman damping is described for hcp metals.

The strength of the interaction, $\mathcal{F}$, between phonons of wave vectors ${q}, {q}', {q}''$ in bands $p, p', p''$, is given in terms of the eigenvalues, $\omega_p({q})$, and eigenvectors, $e_p^{\tau \alpha}({q})$ of the harmonic hamiltonian, as well as the third derivative of the potential energy, $\Phi_{0\tau_1, R_2\tau_2, R_3\tau_3}^{\alpha_1 \alpha_2 \alpha_3}$ \cite{peierls}, 

\scriptsize
\begin{widetext}
\begin{equation}
\mathcal{F} \begin{array}{r} \{p  p'  p''\} \\ \{{q}  {q}'  {q}''\}\end{array}  =\left( \frac{\hbar}{2}\right)^{3/2} \frac{1}{\sqrt{N}}  \sum_{\tau_1, \tau_2,\tau_3}\sum_{ \alpha_1, \alpha_2, \alpha_3} \left( \sum_{R_2, R_3} e^{-i{q}' \cdot {a}_{R_2} -i{q}'' \cdot {a}_{R_3}} \frac{ \Phi_{0\tau_1, R_2\tau_2, R_3\tau_3}^{\alpha_1 \alpha_2 \alpha_3}}{\sqrt{m_{\tau_1}m_{\tau_2}m_{\tau_3}}} \right) \frac{e_{p}^{\tau_1 \alpha_1}({q})e_{p'}^{\tau_2 \alpha_2}({q'})e_{p''}^{\tau_3 \alpha_3}({q''})}{\sqrt{\omega_{p}({q})\omega_{p'}({q'})\omega_{p''}({q''})}}.  \label{eq:2}
\end{equation} 
\end{widetext}
\normalsize
In the above equations, ${a}_i$ are any lattice vectors of a crystal containing $N$ cells, and $\Delta r_{R \tau}^{\alpha}$ a displacement of atom $\tau$ with mass $m_{\tau}$ in cell $R$ in the direction $\alpha$ around the equilibrium position.

According to  the second order many body perturbation theory, the third order hamiltonian $H_3$ produces the self energy $\Sigma_p({q},\omega)=\Lambda_p({q},\omega)+i\Gamma_p({q},\omega)$ with $\Gamma_p({q},\omega)=\sum_{p' p''} \Gamma_{p p' p''}({q},\omega) $ and\\
\begin{equation}
 \Gamma_{p p' p''}({q},\omega)=\frac{\pi}{\hbar^2}\sum_{q'} \bigg| \mathcal{F} \begin{array}{r}  \{p  p'  p''\} \\  \{-{q}  {q}' {q}''\}\end{array}  \bigg|^2 f  \begin{array}{r}  \{p  p'  p''\} \\  \{{q}  {q}' {q}''\}\end{array} .   \label{eq:6}
\end{equation}
The function $f$ is the temperature dependent part given in term of the Bose-Einstein occupation factor $n_{qp}$ by 
\begin{equation*}
f  \begin{array}{r}  \{p  p'  p''\} \\  \{{q}  {q}' {q}''\}\end{array}=(n_{q'p'}-n_{q'' p''})\delta(\omega+\omega_{q'p'}-\omega_{q''p''})+1/2(1+n_{q'p'}+n_{q'' p''})\delta(\omega-\omega_{q'p'}-\omega_{q''p''}) .
\end{equation*} 
However in this paper we are only concerned with the $T=0$ limit where the first term vanishes.

The calculation of the self energy can be greatly improved if we use the symmetry properties of the coupling function $\mathcal{F}$. Let us denote by $\mathcal{P}$ the set of permutation operations $\mathcal{P}=\{ 1, P_{23},P_{12},P_{13},P_{12}P_{23},P_{13}P_{32}\}$ where $P_{ij}$ switch the $i$ and $j$ element of any triplet. For example $P_{12}\{a,b,c \}=\{b,a,c\}$.
The set of rotations of the point group of the crystal is called $\mathcal{R}$. To make the equations more compact when such rotations are applied to a triplet of wave vectors $\{ {q}{q}'{q}'' \}$, we will use the notation $R\{ {q}{q}'{q}'' \}=\{ R {q}R {q}'R {q}'' \} \,\,\, \forall R  \in \mathcal{R}$. The first vector of such a triplet is written as $R\{ {q} |{q}' {q}'' \}$.

From the definition (\ref{eq:2}) of the coupling function $\mathcal{F}$, it is then straightforward to show the following properties, \\
\begin{equation}
 \mathcal{F} \begin{array}{r} P \{p  p' p''\} \\ P \{{q}  {q}' {q}''\}\end{array}  =\mathcal{F}  \begin{array}{r}  \{p  p' p''\} \\ \{{q}  {q}'  {q}''\}\end{array} \,\,\, \forall P  \in \mathcal{P} , \label{eq:3}
\end{equation}
\begin{equation}
 \mathcal{F} \begin{array}{r}  \{p  p'  p''\} \\ P \{{q}  {q}' {q}''\}\end{array}  =\mathcal{F}  \begin{array}{r}  P^{-1}\{p  p'  p''\} \\ \{{q}  {q}'  {q}''\}\end{array}  \,\,\, \forall P  \in \mathcal{P} . \label{eq:4}
\end{equation}
Using the invariance of the potential energy under the space group operations of the crystal and the law of transformation for the eigenvectors\cite{maradudin-1968}, one can also show that,$ \forall R  \in \mathcal{R}$\\
\begin{equation}
 \mathcal{F} \begin{array}{r}  \{p  p'  p''\} \\ R \{{q}  {q}' {q}''\}\end{array} =\mathcal{F}  \begin{array}{r} \{p  p'  p''\} \\ \{{q}  {q}'  {q}''\}\end{array}  \,\,\, \textrm{ if} \,\,\, {q}+{q}'+{q}''={G},  \label{eq:5}
\end{equation}
where ${G}$ is a reciprocal lattice vector.\\
The symmetry properties (\ref{eq:3}) (\ref{eq:4}) and (\ref{eq:5}) can now be used to define a set of irreducible triplets of wave vectors $\{{q} {q}' {q}'' \}$. By definition a set of irreducible triplet is a minimal set of triplets $\{{k} {k}' {k}'' \}$, which sum up to a reciprocal lattice vector and  can be used to generate any triplet $\{{q} {q}' {q}'' \}$, which also sums up to a reciprocal lattice vector,  by application of the elements of $\mathcal{P}  \times \mathcal{R}$. In short $\{{q} {q}' {q}'' \}=RP\{{k} {k}' {k}'' \}$.

According to this definition it is sufficient to calculate the coupling function $\mathcal{F}$ for a set of irreducible triplets since all other can be deduced from it. If $\{{q} {q}' {q}'' \}=RP\{{k} {k}' {k}'' \}$, then one has \\
\begin{equation}
 \mathcal{F} \begin{array}{r}  \{p  p'  p''\} \\  \{{q}  {q}' {q}''\}\end{array}  = \mathcal{F}  \begin{array}{r} P^{-1}\{p  p'  p''\} \\ \{{k}  {k}'  {k}''\}\end{array}.  \\
\end{equation}
By analogy to the reduction of Brillouin zone integration to its irreducible part for properties such as the density of states, formula (\ref{eq:6}) for the self energy is now reduced to a sum over irreducible triplets. Since ${q}''$ is determined from crystal momentum conservation, the summation over ${q}'$ can be seen as a sum over all triplets starting with wave vector ${q}$. 
 Then one can write the contribution to $\Gamma_p({q},\omega)$ for a phonon decaying to bands $p'$ and $p''$ as \\
\begin{widetext}
\begin{eqnarray}
 \Gamma_{p p' p''}(-{q},\omega)&=& \frac{\pi}{\hbar^2}  \sum_{\{ {k} {k}' {k}'' \}} \sum_{P \in \mathcal{P}} \sum_{R \in \mathcal{R}} |  \mathcal{F}  \begin{array}{r}  \{p  p'  p''\} \\  PR\{{k}  {k}' {k}''\}\end{array}  |^2 f \begin{array}{r}  \{p  p'  p''\} \\  PR \{{k}  {k}' {k}''\}\end{array}  \delta_1({q},PR\{{k} | {k}' {k}''\}) \\
&=& \frac{\pi}{\hbar^2}  \sum_{\{ {k} {k}' {k}'' \}} \sum_{P \in \mathcal{P}}  | \mathcal{F}  \begin{array}{r} P^{-1} \{p  p'  p''\} \\ \{{k}  {k}' {k}''\}\end{array}  |^2 f  \begin{array}{r}  \{p  p'  p''\} \\  P \{{k}  {k}' {k}''\}\end{array}  \mathcal{C}_{P}({q},\{{k} {k}' {k}''\}). \label{eq:7}
\end{eqnarray}
\end{widetext}
The $\delta_1$ function is different from zero if ${q}=PR\{{k} | {k}' {k}''\}$ and is given by the reciprocal of the number of times the triplet $PR\{{k}  {k}' {k}''\}$ has been generated by application of the operations of $\mathcal{P} \times \mathcal{R}$ on $\{{k}  {k}' {k}''\}$. The second line is obtained using the symmetry properties (\ref{eq:3}), (\ref{eq:4}), (\ref{eq:5}) and $RP=PR$.
The weight coeficients $\mathcal{C}_P$ are calculated once for all from $\mathcal{C}_P({q},\{{k} {k}' {k}''\})=\sum_{R \in \mathcal{R}}\delta_1({q},PR\{{k} | {k}' {k}''\})$, and it is also usefull to remark that $\mathcal{C}_1=\mathcal{C}_{P_{23}}$,  $\mathcal{C}_{P_{12}}=\mathcal{C}_{P_{13}P_{32}}$ and $\mathcal{C}_{P_{13}}=\mathcal{C}_{P_{12}P_{23}}$.
Equation (\ref{eq:7}) is particularly useful for computer calculations since it can easily be parallelized over irreducible triplets of wave vectors, as with single \emph{k}-point for one particle properties.
A computer code has been implemented from these equations, and in the following it is applied to bcc, fcc and hcp transition metals.

With the exception of manganese, the transition metals crystalize in the bcc, fcc and hcp structures. The second- and third-order force constants can be seen as  derivatives of potential energy or dervatives of forces which are obtained from these structures using first principle calculations. In particular the third order force constants are third derivative of
potential energy with respect to atomic displacements. Therefore
they are calculated from forces on atoms in a supercell containing
two atomic displacements. The total number of atomic displacement pairs is
reduced using crystal symmetry. In our study, we employed finite displacement method to calculate the
derivatives, but to improve the accuracy displacements of plus
and minus directions are applied if they are not symmetrically
equivalent. The third order force constants are usually overdetermined
in this way. The tensor elements are then determined using pseudo inverse,
which is the technique also  employed for second-order force constants
\cite{parlinski-1997}. The details of these calculations are given in the appendix.

To obtain the electronic structure and forces we employed the projector augmented wave method~\cite{PAW-Blochl-1994}, in the framework of density functional theory, within the generalized
gradient approximation of  Perdew, Burke and Ernzerhof \cite{Perdew-PBE-1996} as implemented in the VASP code\cite{VASP-Kresse-1995,VASP-Kresse-1996,VASP-Kresse-1999}. Spin polarized calculations are performed for Fe, Cr, Ni and Co. The supercells of the bcc, fcc and hcp structures contain 16, 32, 16
atoms respectively, and are limited by our
computational resources. A plane-wave energy cutoff of 300 eV is
used, and k-point sampling meshes of $12 \times 12 \times 12$, $12 \times 12 \times 12 $ and $16 \times
16 \times 8$ are used for bcc, fcc and hcp supercells, respectively.  The
Methfessel-Paxton scheme\cite{Methfessel-1989} is employed with a smearing of
0.2 eV. The cell parameters are relaxed using until the stresses becomes less than
10$^{-3}$ GPa. Atomic forces are obtained
with an energy convergence criterion of 10$^{-8}$ eV.  For some metals, such as Cr, the electronic ground state we obtain from density functional theory can be questionable. However the forces extracted from those calculations may still be used to calculate the forces constants. For example for Cr we have checked that the second order forces constants gives a phonons spectrum,  at the point N of the Brillouin zone, at most different by 6 \% of the experimental values \cite{shaw71}. Since the third order force constants is even more short ranged, we assume this approximation to be still acceptable. 

\begin{figure}[ht]
  \begin{center}
   \includegraphics[scale=0.55]{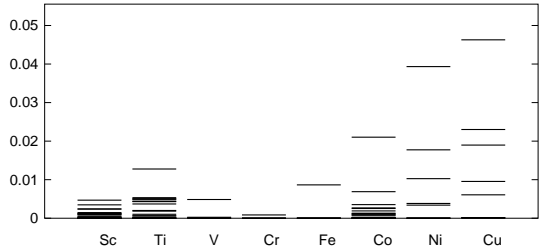}
   \caption{Imaginary part of the self energy in THz. The values are given for all bands and all high symetry points of the Brillouin zones.} \label{fig:1}
  \end{center}
\end{figure}

According to equation (\ref{eq:7}) and the force constants previously calculated we can obtain the damping functions. They are calculated for all bands and all high symmetry points\footnote{points H, N, P for bcc, K, L, W, X for fcc, and $\Gamma$, A, H, K, M, L for hcp.} in the first Brillouin zone. These functions are non zero between $0$ and $2 \omega_{max}$ but we found the center of gravity located at about $2 \bar{\omega}$, where $\bar{\omega}$ is the average phonon frequency over the Brillouin zone. Those are quite smooth functions for bcc and fcc metals whereas they exhibit a more complicated structure for hcp metals.

The probability decays of harmonic phonons are then found according to the equation $1/2 \tau_{qp}=Im \Sigma_{qp}(\omega_{qp})$. They are represented on figure \ref{fig:1} for all bands and all high symmetry points of the Brillouin zones. One can see that the minimum lifetime tend to decrease toward the right end of the series. But even if most of the calculated values are greater than $0.5 \times 10^{-10}$ s, one cannot distinguish a clear trend in their distribution for a given crystal symmetry.

\begin{figure}[ht]
  \begin{center}
   \includegraphics[scale=0.5]{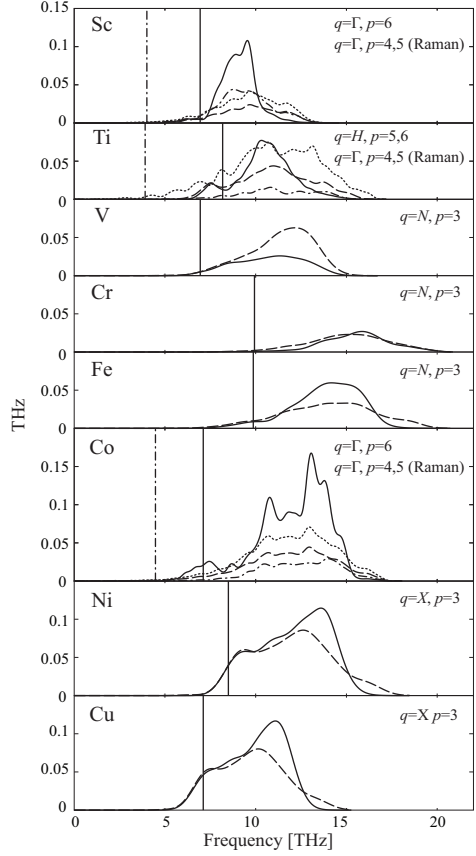}
   \caption{The imaginary part of the self energy calculated at the point $q$ and band $p$ is represented as a functions of frequency. The continuous line is the full calculation and the dashed one correspond to the approximation due to Peierls. For the hcp metals the functions for the Raman active modes are shown with dash-dotted line and their approximation as dotted line. } \label{fig:2}
  \end{center}
\end{figure}

The imaginary parts of the self energy corresponding to the minimum lifetime are represented in figure \ref{fig:2} as functions of frequency. The vertical line shows the frequency of the irreducible representation it belongs to. They are always located within the lower tail of the self energy.\\
\begin{figure*}[ht]
  \begin{center}
   \includegraphics[scale=0.9]{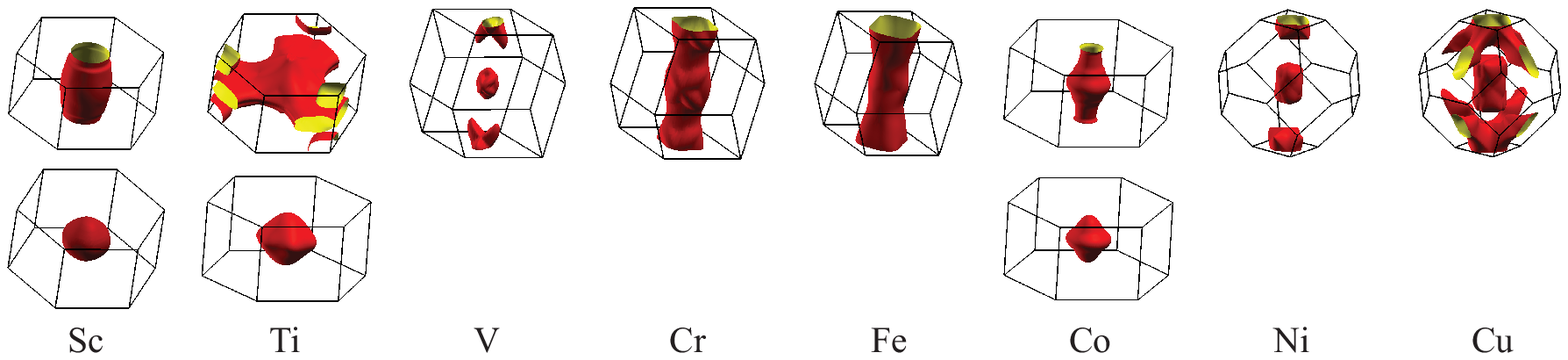}
   \caption{Conservation surfaces for the bands with the stronger probability decays. $q_0$ is chosen as the vertical direction. The surfaces are generated using the Xcrysden software \cite{xcrysden}. The band indices and the contribution to the damping function (in percentages) at the harmonic frequency are as follow : Sc$(p',p''=4,1 ; \,36 \%)$, Sc-Raman$(p',p''=2,1; \,76 \%)$ , Ti $(p',p''=3,1;\, 27 \%)$, Ti-Raman $(p',p''=2,1; 50 \%)$, V $(p',p''=2,1;\, 80 \%)$, Cr $(p',p''=1,1;\, 50 \%)$, Fe $(p',p''=2,1;\, 41 \%)$, Co $(p',p''=4,2;\, 28 \%)$ Co-Raman $(p',p''=2,2;\, 47 \%)$, Ni $(p',p''=2,1;\, 59 \%)$, Cu $(p',p''=2,1;\, 51 \%)$. The three surfaces at the second row correspond to Raman decay.}\label{fig:3}
  \end{center}
\end{figure*}
To better understand how a harmonic phonon of frequency $\omega_{q_0 p_0}$, which will be the phonons with minimum lifetime, acquires a finite lifetime $\Gamma_{q_0 p_0}(\omega_{q_0 p_0})$, one should remember that this quantity is constructed from two parts. The one with the delta functions gives the decay processes which are allowed by the conservation laws and the $\bigg| \mathcal{F}  \begin{array}{r}  \{p  p'  p''\} \\ \{{q}  {q}' {q}''\}\end{array}  \bigg|^2$  gives the probability for such decays to happen. The two conservation laws, for energy $\omega_{q_0 p_0}=\omega_{q' p'}+\omega_{q'' p''}$, and momentum ${q}_0={q}'+{q}''+{G}$, are coupled equations which define a \emph{conservation surface} in reciprocal space: a phonon in mode $q_0 p_0$ will decay in two phonons of bands $p'$ and $p''$ with wave vectors having their extremities on that surface. For each metal at least one couple of bands has a large probability decay. The conservation surfaces corresponding to the strongest ones are plotted in figure $\ref{fig:3}$ and the percentages for such decays are given in the caption. In such  a way one obtains a very clear view of the processes which generates the lifetime since we know the bands to which the phonons decay as well as their wave vectors (the band indices are given in figure \ref{fig:2}). For given $p'$ and $p''$ we should however remember that when $p' \neq p''$, the surface is always composed of two sheets. One centered at the origin where $q'$ is located, and the same shifted by ${q}_0$ where $q''$ lies in. For clarity only the first one is represented on figure $\ref{fig:3}$. All surfaces we found are open surfaces, and with the exception of Ti, they have a tube shape along ${q}_0$.
\newpage
Now if we consider the frequency $\omega_{q_0 p_0}$ as a variable parameter $\omega$, we generate a family of surfaces $S(\omega)$ whose shape and area give the joint density of states, 
\begin{equation*}
D_2(q_0,\omega)=\frac{1}{\Omega} \sum_{p' p''} \int_{S(\omega)} \frac{dS}{\nabla(\omega_{q' p'}+\omega_{q_0-q' p''})}. 
\end{equation*}

As an approximation, Peierls\cite{peierls} proposed the damping function $\Gamma_{q_0 p_0}(\omega)$ to be proportional to the joint density of states. In fact one can also simply fix the proportionality constant replacing $\bigg| \mathcal{F}  \begin{array}{r}  \{p  p'  p''\} \\ \{{q}  {q}' {q}''\}\end{array}  \bigg|^2$ by its average value and rescaling by $\omega_{qp}$,
\begin{eqnarray*}
\left\langle  | \mathcal{F} |^2  \right\rangle  &\sim& \frac{\mathcal{A}}{ \bar{\omega}_{q}\bar{\omega}^2} \xrightarrow{} \frac{\mathcal{A}}{\omega_{qp} \bar{\omega}^2}
\end{eqnarray*}
where we defined $\mathcal{A}$ as $\frac{1}{(3 N_a)^3} \sum_{\tau_1 \tau_2 \tau_3} \sum_{\alpha_1 \alpha_2 \alpha_3} \sum_{R_2 R_3} ( \Phi_{0 \tau_1, R_2\tau_2, R_3\tau_3}^{\alpha_1 \alpha_2 \alpha_3})^2$. It represents an averaged measure of the anharmonicity. $\bar{\omega}_{q}$ is the average value of frequencies at point $q$. \\
The approximation is clearly good for fcc and even bcc transition metals where the anharmonicity seems to be describable by a single number $\alpha$ for these processes. However the approximation is not accurate for hcp metals. The coupling function has a much stronger dependance on the phonon modes. This comes from the hexagonal structure which has two atoms per cell. This combination gives third order force constants which are anisotropic, and complicated interferences between phonons eigenvectors, accoustical and optical, which cannot be removed from the calculation.
We have also calculated the Raman damping functions of the hcp transition metals and the same conclusion is obtained. For this structure it is the E$_{2g}$ mode which is Raman active. These phonons have a much longer lifetime with $2.1 \times 10^{-9}$s for Sc, $5.8 \times 10^{-9}$s for Ti and $1.8 \times 10^{-9}$s for Co. Their conservation surfaces are presented at the second row of figure \ref{fig:3} and the damping functions are shown in figure \ref{fig:2}. For this mode the stronger decay to a couple of bands $p' p''$ is much more selective than in the case of the modes with minimum lifetime. Remarkably, an optical phonon in the Raman active mode E$_{2g}$ will decay into a pair of acoustic phonons in almost all cases. The wave vectors of such phonons are located on the surfaces shown in figure \ref{fig:3}. These surfaces are closed, with very simple shapes. This seems to indicate that simpler models could be constructed for these decay processes.

In conclusion, we have calculated the phonon-phonon self energy of bcc, fcc and hcp transition metals. The decays for the phonons with minimum lifetime were studied and the conservation surfaces calculated. We found that for bcc and fcc metals the imaginary part of the self energy is approximately proportional to the joint density of states whereas this approximation fails in the case of hcp metals. The Raman damping was also examined for these metals and we found that a phonon decay into a pair of acoustic phonons whose wave vectors are located on spherical-like surfaces.

The authors gratefully acknowledge the French Agence Nationale de la Recherche (ANR) for
financial support under contract \# 07-MAPR-0015-04 as well as the
Grants-in-Aid for Scientific Research (A), Scientific Re-
search on Priority Areas (Grant No. 474), and the Global
COE Program, all from MEXT, Japan, and the PIE - Programme Interdisciplinaire Energie of CNRS, France.

\appendix
\section{Computations of second- and third-order force constants}
The potential energy of a phonon system is represented as a function of atomic
positions, $V(\mathbf{r}_{R_1 \tau_1},\ldots,\mathbf{r}_{R_N
\tau_n})$, where $\mathbf{r}_{R \tau}$ is the atomic
position, and $n$ and $N$ are the number of atoms in a unit cell and the
number of unit cells, respectively. $\tau_i$ and $R_i$ are the indices
of atoms in a unit cell and the indices of unit cells.

A force on an atom is the first derivative of the potential energy with
respect to an atomic position,
\begin{equation}
 F^{\alpha}_{R \tau} = -\frac{\partial V }{\partial r^{\alpha}_{R \tau}}.
\end{equation}
$\alpha$, $\beta$, ..., are used for the indices of Cartesian
coordinates. A second-order force constant $\Phi^{\alpha \beta}$ is the second
derivative of the potential energy as function of atomic positions,
\begin{equation}
 \label{eq:second-fc}
 \Phi^{\alpha \beta}_{R_1 \tau_1, R_2 \tau_2} = \frac{\partial^2
 V}{\partial r^{\alpha}_{R_1 \tau_1} \partial r^{\beta}_{R_2 \tau_2}} =
 -\frac{\partial F^{\alpha}_{R_1 \tau_1}}{\partial r^{\beta}_{R_2 \tau_2}},
\end{equation}
and a third-order force constant $\Phi^{\alpha \beta \gamma}$ is the third
derivative of the potential energy as function of atomic positions,
\begin{eqnarray}
 \label{eq:third-fc}
 \Phi^{\alpha \beta \gamma}_{R_1 \tau_1, R_2 \tau_2, R_3 \tau_3} & = & \frac{\partial^3
  V}{\partial r^{\alpha}_{R_1 \tau_1} \partial r^{\beta}_{R_2 \tau_2} \partial
  r^{\gamma}_{R_3 \tau_3}} \nonumber \\
 & = & \frac{\partial \Phi^{\alpha \beta}_{R_1 \tau_1, R_2 \tau_2}}
  {\partial r^{\gamma}_{R_3 \tau_3}} \nonumber \\
 & = & -\frac{\partial^2 F^{\alpha}_{R_1 \tau_1}}{\partial r^{\beta}_{R_2 \tau_2}
  \partial r^{\gamma}_{R_3 \tau_3}}.
\end{eqnarray}
Using finite differences, the derivatives in Eqs.~(\ref{eq:second-fc})
and (\ref{eq:third-fc}) are approximated by
\begin{equation}
 \label{eq:finite-fc2}
   \Phi^{\alpha \beta}_{R_1 \tau_1,R_2 \tau_2} \simeq -\frac{
    F^{\alpha}_{R_1 \tau_1}[{\Delta r^{\beta}_{R_2 \tau_2}}]} {\Delta r^{\beta}_{R_2 \tau_2}},
\end{equation}
and
\begin{eqnarray*}
 \Phi^{\alpha \beta \gamma}_{R_1 \tau_1, R_2 \tau_2, R_3 \tau_3}  &\simeq& 
  \frac{\Delta \Phi^{\alpha \beta}_{R_1 \tau_1, R_2 \tau_2 }[{\Delta
  r^{\gamma}_{R_3 \tau_3}}] }{\Delta
  r^{\gamma}_{R_3 \tau_3} } \\
  & =&   \frac{ \Phi^{\alpha \beta}_{R_1 \tau_1, R_2 \tau_2}[ {\Delta
  r^{\gamma}_{R_3 \tau_3}}] - \Phi^{\alpha \beta}_{R_1 \tau_1, R_2 \tau_2 }}{\Delta
  r^{\gamma}(R_3 \tau_3) } \nonumber \\ 
 & \simeq &  - \frac{ F^{\alpha}_{R_1 \tau_1}[ {\Delta
  r^{\beta}_{R_2 \tau_2}},{\Delta
  r^{\gamma}_{R_3 \tau_3}}]  -
  F^{\alpha}_{R_1 \tau_1}[ {\Delta
  r^{\beta}_{R_2 \tau_2}}]
  }
 {\Delta r^{\beta}_{R_2 \tau_2}
  \Delta r^{\gamma}_{R_3 \tau_3}},
\end{eqnarray*}
respectively. $\Delta r^{\beta}$ and $\Delta r^{\gamma}$ correspond to the finite
atomic displacements. The $\Delta r^{\beta}$ and $\Delta r^{\gamma}$ appering  in the parentheses of forces and force constants mean
that the values are calculated under the displacements.

To compute the second-order force constants, we employed the technique
presented by Parlinski {\it et. al}.\cite{parlinski-1997} and the
third-order force constants are obtained in a similar manner. In the following
sections, the computational details are given.

\subsection{Computation of second-order force constants}
Second-order force constants are computed through the approximation
(\ref{eq:finite-fc2}) with small displacements. For
computational convenience, a second-order force constant tensor for a
pair of atoms, $R_1 \tau_1$ and $R_2 \tau_2$, and an atomic displacement are represented by a $9 \times
1$ matrix $\mathbf{P}$ and a $3 \times 9$ matrix $\mathbf{U}$ given
by
\begin{equation}
   \mathbf{P}(R_2 \tau_2,R_1 \tau_1) = [
  \begin{matrix}
   \Phi^{xx} \,
   \Phi^{xy} \,
   \Phi^{xz} \,
   \Phi^{yx} \,
   \Phi^{yy} \,
   \Phi^{yz} \,
   \Phi^{zx} \,
   \Phi^{zy} \,
   \Phi^{zz}\,
  \end{matrix}
  ]^T
\end{equation}
and
\begin{equation}
 \mathbf{U}(R_2 \tau_2) =    \begin{pmatrix}
   \mathbf{1} & \mathbf{0} & \mathbf{0} \\
   \mathbf{0} & \mathbf{1} & \mathbf{0} \\
   \mathbf{0} & \mathbf{0} & \mathbf{1} \\
  \end{pmatrix} \otimes  [
  \begin{matrix}
   \Delta r^{x} & \Delta r^{y} & \Delta r^{z}
  \end{matrix}],
\end{equation}
respectively. Using these matrices, a force on an atom, which is in the
form of a $1\times 3$ matrix $\mathbf{F}$, is obtained by
\begin{equation}
 \label{eq:f-up}
 \mathbf{F}(R_1 \tau_1) = -\mathbf{U}(R_2 \tau_2)  \mathbf{P}(R_1 \tau_1,R_2 \tau_2).
\end{equation}
Simultaneous equations of different atomic displacements for a pair of
atoms are then combined as
\begin{equation}
 \label{eq:f-up-more}
 \begin{pmatrix}
 \mathbf{F}_1 \\
 \mathbf{F}_2 \\
  \vdots
 \end{pmatrix}
 = -
 \begin{pmatrix}
 \mathbf{U}_1 \\
 \mathbf{U}_2 \\
  \vdots
 \end{pmatrix}
 \mathbf{P}.
\end{equation}
With sufficient number of atomic displacements, Eq.~(\ref{eq:f-up-more})
may be solved by pseudo inverse such as
\begin{equation}
 \label{eq:pinv2}
 \mathbf{P} = -
 \begin{pmatrix}
 \mathbf{U}_1 \\
 \mathbf{U}_2 \\
  \vdots
 \end{pmatrix}^{+}
 \begin{pmatrix}
 \mathbf{F}_1 \\
 \mathbf{F}_2 \\
  \vdots
 \end{pmatrix}.
\end{equation}
However with the help of site-point symmetry, the required number of atomic
displacements to solve the simultaneous equations may be reduced. If $R_1' \tau_1'$ is the image of atom $R_1 \tau_1$ by a
site-point symmetry operation of atom $R_2 \tau_2$, Eq.~(\ref{eq:f-up}) becomes
\begin{eqnarray}
 \mathbf{F}(R_1' \tau_1')&=&-\mathbf{U}(R_2 \tau_2)  \mathbf{P}(R_1' \tau_1',R_2 \tau_2)\\
                                  &=& -\mathbf{U}(R_2 \tau_2) \mathbf{A} \mathbf{P}(R_1 \tau_1,R_2 \tau_2)  \label{eq:uap}
\end{eqnarray}
where $\mathbf{F}(R_1' \tau_1')$ is the force at the atomic site obtained from the
original atomic site by the site-point symmetry operation, and
$\mathbf{A}$ is the $9\times 9$ matrix that is used to rotate
$\mathbf{P}$ along the site-point symmetry operation.  Using
Eq.~(\ref{eq:uap}), the combined simultaneous equations are built such
as
\begin{equation}
 \label{eq:uap-combine}
 \begin{pmatrix}
 \mathbf{F}^{(1)}_1 \\
 \mathbf{F}^{(2)}_1 \\
  \vdots \\
 \mathbf{F}^{(1)}_2 \\
 \mathbf{F}^{(2)}_2 \\
  \vdots
 \end{pmatrix}
 = -
 \begin{pmatrix}
 \mathbf{U}_1  \mathbf{A}^{(1)} \\
 \mathbf{U}_1  \mathbf{A}^{(2)} \\
  \vdots \\
 \mathbf{U}_2  \mathbf{A}^{(1)} \\
 \mathbf{U}_2  \mathbf{A}^{(2)} \\
  \vdots
 \end{pmatrix}
 \mathbf{P}.
\end{equation}
where the superscript with parenthesis gives the symmetry operation
index. This is solved like Eq.~(\ref{eq:pinv2}).

\subsection{Computatation of third-order force constants}

The finite difference approximation for the third-order force constants is represented by matrices as
\begin{equation}
 \Delta \mathbf{P}(R_1 \tau_1,R_2 \tau_2) = \mathbf{V}(R_3 \tau_3) \cdot \mathbf{Q}(R_1 \tau_1, R_2 \tau_2, R_3 \tau_3)
\end{equation}
where $\Delta \mathbf{P}$, $\mathbf{V}$, and $\mathbf{Q}$ are the $9\times 1$, $9\times 27$ and $27 \times 1$
matrices corresponding to $\Delta\Phi^{\alpha \beta}$, $\Delta r^{\gamma}$, and
$\Phi^{\alpha \beta \gamma}$, respectively, and are given by

\begin{equation}
 \Delta\mathbf{P}_{\beta+3(\alpha-1)}= \Delta \Phi^{\alpha \beta}  \,\,\,\,\,\,\, \alpha,\beta= 1,2,3
\end{equation}

\begin{equation}
 \mathbf{Q}_{\gamma+3(\beta-1)+9(\alpha-1)}=\Phi^{\alpha \beta \gamma} \,\,\,\,\,\,\, \alpha,\beta,\gamma= 1,2,3
\end{equation}
and
\begin{equation}
 \mathbf{V} =
  \begin{pmatrix}
   \mathbf{1} & \mathbf{0} & \mathbf{0} \\
   \mathbf{0} & \mathbf{1} & \mathbf{0} \\
   \mathbf{0} & \mathbf{0} & \mathbf{1} \\
  \end{pmatrix}\otimes  \mathbf{U} ,
\end{equation}
respectively. Simultaneous equations are constructed by a similar manner
to Eq.~(\ref{eq:f-up-more}) as
\begin{equation}
  \begin{pmatrix}
   \Delta\mathbf{P_1}\\
   \Delta\mathbf{P_2}\\
   \vdots
  \end{pmatrix}
  =
  \begin{pmatrix}
   \mathbf{V_1}\\
   \mathbf{V_2}\\
   \vdots
  \end{pmatrix}
  \mathbf{Q}.
\end{equation}
This may solved by pseudo inverse such as
\begin{equation}
 \label{eq:pinv3}
 \mathbf{Q} =
  \begin{pmatrix}
   \mathbf{V_1}\\
   \mathbf{V_2}\\
   \vdots
  \end{pmatrix}^{+}
  \begin{pmatrix}
   \Delta\mathbf{P_1}\\
   \Delta\mathbf{P_2}\\
   \vdots
  \end{pmatrix}.
\end{equation}
The number of pair of displacements to calculate can be reduced using symmetry operations that conserve a third-order force constant
tensor for a triplet of atoms. If  $R_1' \tau_1'$ is the image of atom $R_1 \tau_1$ through a symmetry of the displaced structure, then one has
\begin{eqnarray*}
\Delta \mathbf{P}(R_1' \tau_1',R_2 \tau_2) &=& \mathbf{V}(R_3 \tau_3) \cdot \mathbf{Q}(R_1' \tau_1',R_2 \tau_2,R_3 \tau_3) \\
                                                         &=& \mathbf{V}(R_3 \tau_3) \cdot  \mathbf{B} \cdot \mathbf{Q}(R_1 \tau_1,R_2 \tau_2,R_3 \tau_3) 
\end{eqnarray*}
where  $\mathbf{B}$  is the $27\times 27$ symmetry operation matrix that
transform the tensor $\mathbf{Q}$.\\
The  simultaneous equations are then written in a
similar manner to Eq.~(\ref{eq:uap-combine}) as
\begin{equation}
  \begin{pmatrix}
   \Delta\mathbf{P}^{(1)}_{1}\\
   \Delta\mathbf{P}^{(2)}_{1}\\
   \vdots\\
   \Delta\mathbf{P}^{(1)}_{2}\\
   \Delta\mathbf{P}^{(2)}_{2}\\
   \vdots
  \end{pmatrix}=
  \begin{pmatrix}
   \mathbf{V}_1 \mathbf{B}^{(1)} \\
   \mathbf{V}_1 \mathbf{B}^{(2)} \\
   \vdots\\
   \mathbf{V}_2 \mathbf{B}^{(1)} \\
   \mathbf{V}_2 \mathbf{B}^{(2)} \\
   \vdots
  \end{pmatrix}
  \mathbf{Q}
\end{equation}
and this is solved in the same way as Eq.~(\ref{eq:pinv3}) using the pseudo inverse method.

\bibliography{p-p-tm}

\begin{thebibliography}{18}
\expandafter\ifx\csname natexlab\endcsname\relax\def\natexlab#1{#1}\fi
\expandafter\ifx\csname bibnamefont\endcsname\relax
  \def\bibnamefont#1{#1}\fi
\expandafter\ifx\csname bibfnamefont\endcsname\relax
  \def\bibfnamefont#1{#1}\fi
\expandafter\ifx\csname citenamefont\endcsname\relax
  \def\citenamefont#1{#1}\fi
\expandafter\ifx\csname url\endcsname\relax
  \def\url#1{\texttt{#1}}\fi
\expandafter\ifx\csname urlprefix\endcsname\relax\def\urlprefix{URL }\fi
\providecommand{\bibinfo}[2]{#2}
\providecommand{\eprint}[2][]{\url{#2}}

\bibitem[{\citenamefont{Baroni et~al.}(1987)\citenamefont{Baroni, Giannozzi,
  and Testa}}]{baroni-1987}
\bibinfo{author}{\bibfnamefont{S.}~\bibnamefont{Baroni}},
  \bibinfo{author}{\bibfnamefont{P.}~\bibnamefont{Giannozzi}},
  \bibnamefont{and} \bibinfo{author}{\bibfnamefont{A.}~\bibnamefont{Testa}},
  \bibinfo{journal}{Phys. Rev. Lett.} \textbf{\bibinfo{volume}{58}},
  \bibinfo{pages}{1861} (\bibinfo{year}{1987}).

\bibitem[{\citenamefont{Parlinski et~al.}(1997)\citenamefont{Parlinski, Li, and
  Kawazoe}}]{parlinski-1997}
\bibinfo{author}{\bibfnamefont{K.}~\bibnamefont{Parlinski}},
  \bibinfo{author}{\bibfnamefont{Z.~Q.} \bibnamefont{Li}}, \bibnamefont{and}
  \bibinfo{author}{\bibfnamefont{Y.}~\bibnamefont{Kawazoe}},
  \bibinfo{journal}{Phys. Rev. Lett.} \textbf{\bibinfo{volume}{78}},
  \bibinfo{pages}{4063} (\bibinfo{year}{1997}).

\bibitem[{\citenamefont{Togo et~al.}(2008)\citenamefont{Togo, Oba, and
  Tanaka}}]{phonopy}
\bibinfo{author}{\bibfnamefont{A.}~\bibnamefont{Togo}},
  \bibinfo{author}{\bibfnamefont{F.}~\bibnamefont{Oba}}, \bibnamefont{and}
  \bibinfo{author}{\bibfnamefont{I.}~\bibnamefont{Tanaka}},
  \bibinfo{journal}{Phys. rev. B} \textbf{\bibinfo{volume}{78}},
  \bibinfo{pages}{134106} (\bibinfo{year}{2008}).

\bibitem[{\citenamefont{Pavone et~al.}(1993)\citenamefont{Pavone, Karch,
  SchŸtt, Strauch, and Windl}}]{pavone}
\bibinfo{author}{\bibfnamefont{P.}~\bibnamefont{Pavone}},
  \bibinfo{author}{\bibfnamefont{K.}~\bibnamefont{Karch}},
  \bibinfo{author}{\bibfnamefont{O.}~\bibnamefont{SchŸtt}},
  \bibinfo{author}{\bibfnamefont{D.}~\bibnamefont{Strauch}}, \bibnamefont{and}
  \bibinfo{author}{\bibfnamefont{W.}~\bibnamefont{Windl}},
  \bibinfo{journal}{Phys. rev. B} \textbf{\bibinfo{volume}{48}},
  \bibinfo{pages}{3156} (\bibinfo{year}{1993}).

\bibitem[{\citenamefont{Togo et~al.}(2010)\citenamefont{Togo, Chaput, Tanaka,
  and Hug}}]{togo10}
\bibinfo{author}{\bibfnamefont{A.}~\bibnamefont{Togo}},
  \bibinfo{author}{\bibfnamefont{L.}~\bibnamefont{Chaput}},
  \bibinfo{author}{\bibfnamefont{I.}~\bibnamefont{Tanaka}}, \bibnamefont{and}
  \bibinfo{author}{\bibfnamefont{G.}~\bibnamefont{Hug}},
  \bibinfo{journal}{Phys. rev. B} \textbf{\bibinfo{volume}{81}},
  \bibinfo{pages}{174301} (\bibinfo{year}{2010}).

\bibitem[{\citenamefont{Deinzer et~al.}(2003)\citenamefont{Deinzer, Birner, and
  Strauch}}]{deinzer-2003}
\bibinfo{author}{\bibfnamefont{G.}~\bibnamefont{Deinzer}},
  \bibinfo{author}{\bibfnamefont{G.}~\bibnamefont{Birner}}, \bibnamefont{and}
  \bibinfo{author}{\bibfnamefont{D.}~\bibnamefont{Strauch}},
  \bibinfo{journal}{Phys. Rev. B} \textbf{\bibinfo{volume}{67}},
  \bibinfo{pages}{144304} (\bibinfo{year}{2003}).

\bibitem[{\citenamefont{Narasimhan and Vanderbilt}(1991)}]{narasimhan-1991}
\bibinfo{author}{\bibfnamefont{S.}~\bibnamefont{Narasimhan}} \bibnamefont{and}
  \bibinfo{author}{\bibfnamefont{D.}~\bibnamefont{Vanderbilt}},
  \bibinfo{journal}{Phys. Rev. B} \textbf{\bibinfo{volume}{43}},
  \bibinfo{pages}{4541} (\bibinfo{year}{1991}).

\bibitem[{\citenamefont{Bonini et~al.}(2007)\citenamefont{Bonini, Lazzeri,
  Marzari, and Mauri}}]{bonini-2007}
\bibinfo{author}{\bibfnamefont{M.}~\bibnamefont{Bonini}},
  \bibinfo{author}{\bibfnamefont{M.}~\bibnamefont{Lazzeri}},
  \bibinfo{author}{\bibfnamefont{N.}~\bibnamefont{Marzari}}, \bibnamefont{and}
  \bibinfo{author}{\bibfnamefont{F.}~\bibnamefont{Mauri}},
  \bibinfo{journal}{Phys. Rev. Lett.} \textbf{\bibinfo{volume}{99}},
  \bibinfo{pages}{176802} (\bibinfo{year}{2007}).

\bibitem[{\citenamefont{Peierls}(1964)}]{peierls}
\bibinfo{author}{\bibfnamefont{R.~E.} \bibnamefont{Peierls}},
  \emph{\bibinfo{title}{Quantum Theory of Solids}} (\bibinfo{publisher}{Oxford
  University Press}, \bibinfo{year}{1964}).

\bibitem[{\citenamefont{Maradudin and Vosko}(1968)}]{maradudin-1968}
\bibinfo{author}{\bibfnamefont{A.~A.} \bibnamefont{Maradudin}}
  \bibnamefont{and} \bibinfo{author}{\bibfnamefont{S.~H.} \bibnamefont{Vosko}},
  \bibinfo{journal}{Rev. Mod. Phys.} \textbf{\bibinfo{volume}{40}},
  \bibinfo{pages}{1} (\bibinfo{year}{1968}).

\bibitem[{\citenamefont{Bl\"{o}chl}(1994)}]{PAW-Blochl-1994}
\bibinfo{author}{\bibfnamefont{P.~E.} \bibnamefont{Bl\"{o}chl}},
  \bibinfo{journal}{Phys. Rev. B} \textbf{\bibinfo{volume}{50}},
  \bibinfo{pages}{17953} (\bibinfo{year}{1994}).

\bibitem[{\citenamefont{Perdew et~al.}(1996)\citenamefont{Perdew, Burke, and
  Ernzerhof}}]{Perdew-PBE-1996}
\bibinfo{author}{\bibfnamefont{J.~P.} \bibnamefont{Perdew}},
  \bibinfo{author}{\bibfnamefont{K.}~\bibnamefont{Burke}}, \bibnamefont{and}
  \bibinfo{author}{\bibfnamefont{M.}~\bibnamefont{Ernzerhof}},
  \bibinfo{journal}{Phys. Rev. Lett.} \textbf{\bibinfo{volume}{77}},
  \bibinfo{pages}{3865} (\bibinfo{year}{1996}).

\bibitem[{\citenamefont{Kresse}(1995)}]{VASP-Kresse-1995}
\bibinfo{author}{\bibfnamefont{G.}~\bibnamefont{Kresse}}, \bibinfo{journal}{J.
  Non-Cryst. Solids} \textbf{\bibinfo{volume}{193}}, \bibinfo{pages}{222}
  (\bibinfo{year}{1995}).

\bibitem[{\citenamefont{Kresse and Furthm\"{u}ller}(1996)}]{VASP-Kresse-1996}
\bibinfo{author}{\bibfnamefont{G.}~\bibnamefont{Kresse}} \bibnamefont{and}
  \bibinfo{author}{\bibfnamefont{J.}~\bibnamefont{Furthm\"{u}ller}},
  \bibinfo{journal}{Comput. Mater. Sci.} \textbf{\bibinfo{volume}{6}},
  \bibinfo{pages}{15} (\bibinfo{year}{1996}).

\bibitem[{\citenamefont{Kresse and Joubert}(1999)}]{VASP-Kresse-1999}
\bibinfo{author}{\bibfnamefont{G.}~\bibnamefont{Kresse}} \bibnamefont{and}
  \bibinfo{author}{\bibfnamefont{D.}~\bibnamefont{Joubert}},
  \bibinfo{journal}{Phys. Rev. B} \textbf{\bibinfo{volume}{59}},
  \bibinfo{pages}{1758} (\bibinfo{year}{1999}).

\bibitem[{\citenamefont{Methfessel and Paxton}(1989)}]{Methfessel-1989}
\bibinfo{author}{\bibfnamefont{M.}~\bibnamefont{Methfessel}} \bibnamefont{and}
  \bibinfo{author}{\bibfnamefont{A.~T.} \bibnamefont{Paxton}},
  \bibinfo{journal}{Phys. Rev. B} \textbf{\bibinfo{volume}{40}},
  \bibinfo{pages}{3616} (\bibinfo{year}{1989}).

\bibitem[{\citenamefont{Shaw and Muhlestein}(1971)}]{shaw71}
\bibinfo{author}{\bibfnamefont{W.~M.} \bibnamefont{Shaw}} \bibnamefont{and}
  \bibinfo{author}{\bibfnamefont{L.~D.} \bibnamefont{Muhlestein}},
  \bibinfo{journal}{Phys. Rev. B} \textbf{\bibinfo{volume}{4}},
  \bibinfo{pages}{969} (\bibinfo{year}{1971}).

\bibitem[{\citenamefont{Kokalj}(2003)}]{xcrysden}
\bibinfo{author}{\bibfnamefont{A.}~\bibnamefont{Kokalj}},
  \bibinfo{journal}{Comp. Mater. Sci.} \textbf{\bibinfo{volume}{28}},
  \bibinfo{pages}{155} (\bibinfo{year}{2003}).

\end{thebibliography}
\end{document}